\documentstyle[twoside,epsfig]{article}

\catcode`\@=11
\long\def\@makefntext#1{
\protect\noindent \hbox to 3.2pt {\hskip-.9pt  
$^{{\eightrm\@thefnmark}}$\hfil}#1\hfill}		%CAN BE USED 

\def\thefootnote{\fnsymbol{footnote}}
\def\@makefnmark{\hbox to 0pt{$^{\@thefnmark}$\hss}}	%ORIGINAL 
	
\def\ps@myheadings{\let\@mkboth\@gobbletwo
\def\@oddhead{\hbox{}
\rightmark\hfil\eightrm\thepage}   
\def\@oddfoot{}\def\@evenhead{\eightrm\thepage\hfil
\leftmark\hbox{}}\def\@evenfoot{}
\def\sectionmark##1{}\def\subsectionmark##1{}}

\oddsidemargin=\evensidemargin
\renewcommand{\thefootnote}{\fnsymbol{footnote}}

\newcounter{sectionc}\newcounter{subsectionc}\newcounter{subsubsectionc}
\renewcommand{\section}[1] {\vspace{12pt}\addtocounter{sectionc}{1} 
\setcounter{subsectionc}{0}\setcounter{subsubsectionc}{0}\noindent 
	{\tenbf\thesectionc. #1}\par\vspace{5pt}}
\renewcommand{\subsection}[1] {\vspace{12pt}\addtocounter{subsectionc}{1} 
	\setcounter{subsubsectionc}{0}\noindent 
	{\bf\thesectionc.\thesubsectionc. {\kern1pt \bfit #1}}\par\vspace{5pt}}
\renewcommand{\subsubsection}[1] {\vspace{12pt}\addtocounter{subsubsectionc}{1}
	\noindent{\tenrm\thesectionc.\thesubsectionc.\thesubsubsectionc.
	{\kern1pt \tenit #1}}\par\vspace{5pt}}
\newcommand{\nonumsection}[1] {\vspace{12pt}\noindent{\tenbf #1}
	\par\vspace{5pt}}

\newcounter{appendixc}
\newcounter{subappendixc}[appendixc]
\newcounter{subsubappendixc}[subappendixc]
\renewcommand{\thesubappendixc}{\Alph{appendixc}.\arabic{subappendixc}}
\renewcommand{\thesubsubappendixc}
	{\Alph{appendixc}.\arabic{subappendixc}.\arabic{subsubappendixc}}

\renewcommand{\appendix}[1] {\vspace{12pt}
        \refstepcounter{appendixc}
        \setcounter{figure}{0}
        \setcounter{table}{0}
        \setcounter{lemma}{0}
        \setcounter{theorem}{0}
        \setcounter{corollary}{0}
        \setcounter{definition}{0}
        \setcounter{equation}{0}
        \renewcommand{\thefigure}{\Alph{appendixc}.\arabic{figure}}
        \renewcommand{\thetable}{\Alph{appendixc}.\arabic{table}}
        \renewcommand{\theappendixc}{\Alph{appendixc}}
        \renewcommand{\thelemma}{\Alph{appendixc}.\arabic{lemma}}
        \renewcommand{\thetheorem}{\Alph{appendixc}.\arabic{theorem}}
        \renewcommand{\thedefinition}{\Alph{appendixc}.\arabic{definition}}
        \renewcommand{\thecorollary}{\Alph{appendixc}.\arabic{corollary}}
        \renewcommand{\theequation}{\Alph{appendixc}.\arabic{equation}}
        \noindent{\tenbf Appendix \theappendixc #1}\par\vspace{5pt}}
\newcommand{\subappendix}[1] {\vspace{12pt}
        \refstepcounter{subappendixc}
        \noindent{\bf Appendix \thesubappendixc. {\kern1pt \bfit #1}}
	\par\vspace{5pt}}
\newcommand{\subsubappendix}[1] {\vspace{12pt}
        \refstepcounter{subsubappendixc}
        \noindent{\rm Appendix \thesubsubappendixc. {\kern1pt \tenit #1}}
	\par\vspace{5pt}}

\newcommand{\textlineskip}{\baselineskip=13pt}
\newcommand{\smalllineskip}{\baselineskip=10pt}

\def\eightcirc{
\begin{picture}(0,0)
\put(4.4,1.8){\circle{6.5}}
\end{picture}}
\def\eightcopyright{\eightcirc\kern2.7pt\hbox{\eightrm c}} 

\newcommand{\copyrightheading}[1]
	{\vspace*{-2.5cm}\smalllineskip{\flushright
	{\footnotesize FERMILAB-CONF-00/305-E}\\
	 }}

\def\abstracts#1#2#3{{
	\centering{\begin{minipage}{4.5in}\baselineskip=10pt\footnotesize
	\parindent=0pt #1\par 
	\parindent=15pt #2\par
	\parindent=15pt #3
	\end{minipage}}\par}} 

\newcommand{\bibit}{\nineit}

\renewenvironment{thebibliography}[1]
	{\frenchspacing
	 \ninerm\baselineskip=11pt
	 \begin{list}{\arabic{enumi}.}
	{\usecounter{enumi}\setlength{\parsep}{0pt}
	 \setlength{\leftmargin 12.7pt}{\rightmargin 0pt} %FOR 1--9 ITEMS
	 \setlength{\itemsep}{0pt} \settowidth
	{\labelwidth}{#1.}\sloppy}}{\end{list}}

\newcounter{itemlistc}
\newcounter{romanlistc}
\newcounter{alphlistc}
\newcounter{arabiclistc}

\newcommand{\fcaption}[1]{
        \refstepcounter{figure}
        \setbox\@tempboxa = \hbox{\footnotesize Fig.~\thefigure. #1}
        \ifdim \wd\@tempboxa > 5in
           {\begin{center}
        \parbox{5in}{\footnotesize\smalllineskip Fig.~\thefigure. #1}
            \end{center}}
        \else
             {\begin{center}
             {\footnotesize Fig.~\thefigure. #1}
              \end{center}}
        \fi}

\newcommand{\tcaption}[1]{
        \refstepcounter{table}
        \setbox\@tempboxa = \hbox{\footnotesize Table~\thetable. #1}
        \ifdim \wd\@tempboxa > 5in
           {\begin{center}
        \parbox{5in}{\footnotesize\smalllineskip Table~\thetable. #1}
            \end{center}}
        \else
             {\begin{center}
             {\footnotesize Table~\thetable. #1}
              \end{center}}
        \fi}

\def\@citex[#1]#2{\if@filesw\immediate\write\@auxout
	{\string\citation{#2}}\fi
\def\@citea{}\@cite{\@for\@citeb:=#2\do
	{\@citea\def\@citea{,}\@ifundefined
	{b@\@citeb}{{\bf ?}\@warning
	{Citation `\@citeb' on page \thepage \space undefined}}
	{\csname b@\@citeb\endcsname}}}{#1}}

\newif\if@cghi
\def\cite{\@cghitrue\@ifnextchar [{\@tempswatrue
	\@citex}{\@tempswafalse\@citex[]}}
\def\citelow{\@cghifalse\@ifnextchar [{\@tempswatrue
	\@citex}{\@tempswafalse\@citex[]}}
\def\@cite#1#2{{$\null^{#1}$\if@tempswa\typeout
	{IJCGA warning: optional citation argument 
	ignored: `#2'} \fi}}

\def\pmb#1{\setbox0=\hbox{#1}
	\kern-.025em\copy0\kern-\wd0
	\kern.05em\copy0\kern-\wd0
	\kern-.025em\raise.0433em\box0}

\def\fnm#1{$^{\mbox{\scriptsize #1}}$}
\def\fnt#1#2{\footnotetext{\kern-.3em
	{$^{\mbox{\scriptsize #1}}$}{#2}}}

\def\fpage#1{\begingroup
\voffset=.3in
\thispagestyle{empty}\begin{table}[b]\centerline{\footnotesize #1}
	\end{table}\endgroup}

\def\runninghead#1#2{\pagestyle{myheadings}
\markboth{{\protect\footnotesize\it{\quad #1}}\hfill}
{\hfill{\protect\footnotesize\it{#2\quad}}}}
\font\tenrm=cmr10
\font\tenit=cmti10 
\font\tenbf=cmbx10
\font\bfit=cmbxti10 at 10pt
\font\ninerm=cmr9
\font\nineit=cmti9

\font\eightrm=cmr8

\def\qed{\hbox{${\vcenter{\vbox{			%HOLLOW SQUARE
   \hrule height 0.4pt\hbox{\vrule width 0.4pt height 6pt
   \kern5pt\vrule width 0.4pt}\hrule height 0.4pt}}}$}}

\renewcommand{\thefootnote}{\fnsymbol{footnote}}	%USE SYMBOLIC FOOTNOTE

\begin{document}

\runninghead{\boldmath   
Measurement of $\Delta m_d$ Using
a Probability Based Same-Side Tagger at CDF
} {\boldmath 
Measurement of $\Delta m_d$ Using
a Probability Based Same-Side Tagger at CDF
%Measurement of $\Delta m_d$ 
%using a probability based Same-Side Tagger 
%at CDF$\ldots$
}

\normalsize\textlineskip
\thispagestyle{empty}
\setcounter{page}{1}

\copyrightheading{}			%{Vol. 0, No. 0 (1993) 000--000}

\vspace*{0.88truein}

\fpage{1}
\centerline{\bf MEASUREMENT OF $\Delta m_d$ USING A PROBABILITY BASED}
\vspace*{0.035truein}
\centerline{\bf SAME-SIDE TAGGER APPLIED TO LEPTON$+$VERTEX EVENTS}
%\vspace*{0.035truein}
% \centerline{\bf 
%EVENTS AT CDF}
\vspace*{0.37truein}
\centerline{\footnotesize G. BAUER
%\footnote{email: bauerg@fnal.gov
% 10 pt Times Roman, uppercase. Use the footnote to indicate the
% present or permanent address of the author.
% }
}
\centerline{\footnotesize (Representing the CDF Collaboration)
}
\vspace*{0.015truein}
\centerline{\footnotesize\it 
Laboratory for Nuclear Science,
Massachusetts Institute of Technology}
\baselineskip=10pt
\centerline{\footnotesize\it Cambridge, Massachusetts, 02139, USA
}
% \vspace*{10pt}
% \centerline{\footnotesize SECOND AUTHOR}
% \vspace*{0.015truein}
% \centerline{\footnotesize\it Group, Laboratory, Address}
% \baselineskip=10pt
% \centerline{\footnotesize\it City, State ZIP/Zone, Country}
%\vspace*{0.225truein}
%\publisher{(received date)}{(revised date)}

\vspace*{0.21truein}
\abstracts
{A measurement of $\Delta m_d$ is performed using
inclusive lepton+vertex events at CDF.
A probability based Same-Side Tagger was developed 
to tag the initial $b$-flavor of the $B^0_d$,
which suppresses tagging on $B$-decay products.
We find  $\Delta m_d = 0.42 \pm 0.09 \pm 0.03$ ps$^{-1}.$
}{}{}

%\textlineskip			%) USE THIS MEASUREMENT WHEN THERE IS
\vspace*{12pt}			%) NO SECTION HEADING

%\vspace*{1pt}\textlineskip	%) USE THIS MEASUREMENT WHEN THERE IS
%\section{General Appearance}	%) A SECTION HEADING

\vspace*{1pt}\textlineskip	%) USE THIS MEASUREMENT WHEN THERE IS
%\section{Introduction  }	%) A SECTION HEADING
\vspace*{-0.5pt}
%%\noindent
%The\ldots

%\vspace*{1pt}\textlineskip	%) USE THIS MEASUREMENT WHEN THERE IS
%\section{General Appearance}	%) A SECTION HEADING
%\vspace*{-0.5pt}
%\noindent
%Contributions to the {\it International Journal of Modern
%Physics A} will be reproduced by photographing the author's
%submitted typeset manuscript.

\textheight=7.8truein
\setcounter{footnote}{0}
\renewcommand{\thefootnote}{\alph{footnote}}

Measurement of the $B^0$ oscillation frequency $\Delta m$, 
as well as $CP$-violation tests,
critically depend on determining both the 
initial and decay ``b-flavor'' of the meson,
with the initial flavor tag usually
the principle difficulty.
We report a new
$\Delta m_d$ 
measurement\raisebox{-0.1ex}{\cite{Tushar}}
using a modified version of ``Same-Side'' flavor 
tagging %\raisebox{-0.1ex}{\cite{SST}}
adapted to the challenging problem
of tagging an inclusive lepton$+$vertex $B$ sample.

The data ($\sim \! 100$ pb$^{-1}$) are from the 1992-6 Tevatron run.
$B$-selection mimics an earlier
analysis\raisebox{-0.1ex}{\cite{Sample}}, and 
% and is only summarized. It 
consists of $e$ and $\mu$ triggers with 
$p_t(\ell) \!>\! 6$ GeV. Tracks are clustered into jets.
Jet tracks within 
$\Delta R = \!\sqrt{(\Delta\eta)^2 + (\Delta \phi)^2)}\!<\!0.7$ 
of the lepton  
are tried for vertexing
if their impact parameter 
to the primary vertex  is $>2\sigma$.
Secondary vertices pass into our
sample if the lepton is in the vertex,
and it has a transverse decay length $> 2.5$ mm.
This results in 59,881 $e$ and 63,674 $\mu$ events.

This sample has quite high $b$-purity, with 
contaminants determined as follows:
the $c\bar{c}$ component by 
the secondary vertex mass distribution
($ 4\pm 1$\% for $e$, $8\pm2$\% for $\mu$);
electron conversions from $dE/dx$ ($0.8\pm 0.1$\%);
fake $e$ fraction also from $dE/dx$ ($0.4\pm 0.2$\%);
and the fake $\mu$ fraction is determined from the
final $\Delta m_d$ fit\fnm{a}\fnt{a}{The sensitivity
to the fake $\mu$-fraction basically arises from the difference 
in the apparent tagging dilutions observed in the $e$ and $\mu$
samples.} 
($4\pm6$\%).

The initial flavor tag used is a variant 
of our previous Same-Side Tagging (SST).\raisebox{-0.1ex}{\cite{SST}}
The idea\raisebox{-0.1ex}{\cite{Rosner}} is
that the {\it flavor} of a $B$ meson is correlated 
to the {\it charge} of a nearby particle.
This may be due to: a) $B^{**}$ decay, or b) fragmentation
where the type of $B$ meson (determined by the
light quark) leaves a corresponding light antiquark
nearby whose type determines the sign of the $\pi^\pm$
formed.
These process will produce $B^0 \pi^+$ pairs,
and not $B^0 \pi^-$'s.
This type of correlation was first seen 
by OPAL\raisebox{-0.1ex}{\cite{OPAL}} in 
$e^+e^- \!\rightarrow Z^0 \rightarrow b\bar{b}$,
and also in fixed target hadroproduction 
of charm.\raisebox{-0.1ex}{\cite{Summers}}

A specific SST tagging algorithm (``$p_T^{rel}$'')
was used by CDF to measure $\Delta m_d$
in a nearly exclusive $B \rightarrow \ell D^{(*)}X$ 
sample,\raisebox{-0.1ex}{\cite{SST}}
as well as in our 
$\sin(2\beta)$ measurement.\raisebox{-0.1ex}{\cite{sin2beta}}
The issue here is: can SST be used in an inclusive
lepton$+$vertex sample given the significant danger of selecting
a charge correlated $B$-decay product as the tag?

The $p_T^{rel}$-algorithm considered tracks 
with $p_t \!>\! 400$ MeV,
to be within $\Delta R<0.7$ of the $B$,
have an impact parameter to the primary vertex
$<3\sigma$ in the transverse plane.
From this set of tracks, a single track was selected as the tag
based on the minimum $p_T^{rel}$.\raisebox{-0.1ex}{\cite{SST}}
This last criteria is prone to select $B$ daughters,
thus instead of the  $p_T^{rel}$ selection we take 
all accepted tracks and impose the additional probability 
cut on the track to not be a $B$-daughter:
${\cal P}_B(r,\Delta R) < 0.3$.
In this case {\it more} than one track may be selected, 
and the tag sign is the {\it sum charge of all accepted 
tracks}.

\begin{figure}[tbp]
% \vspace*{13pt}
\centerline{
  \epsfxsize=15pc 
  \epsfbox{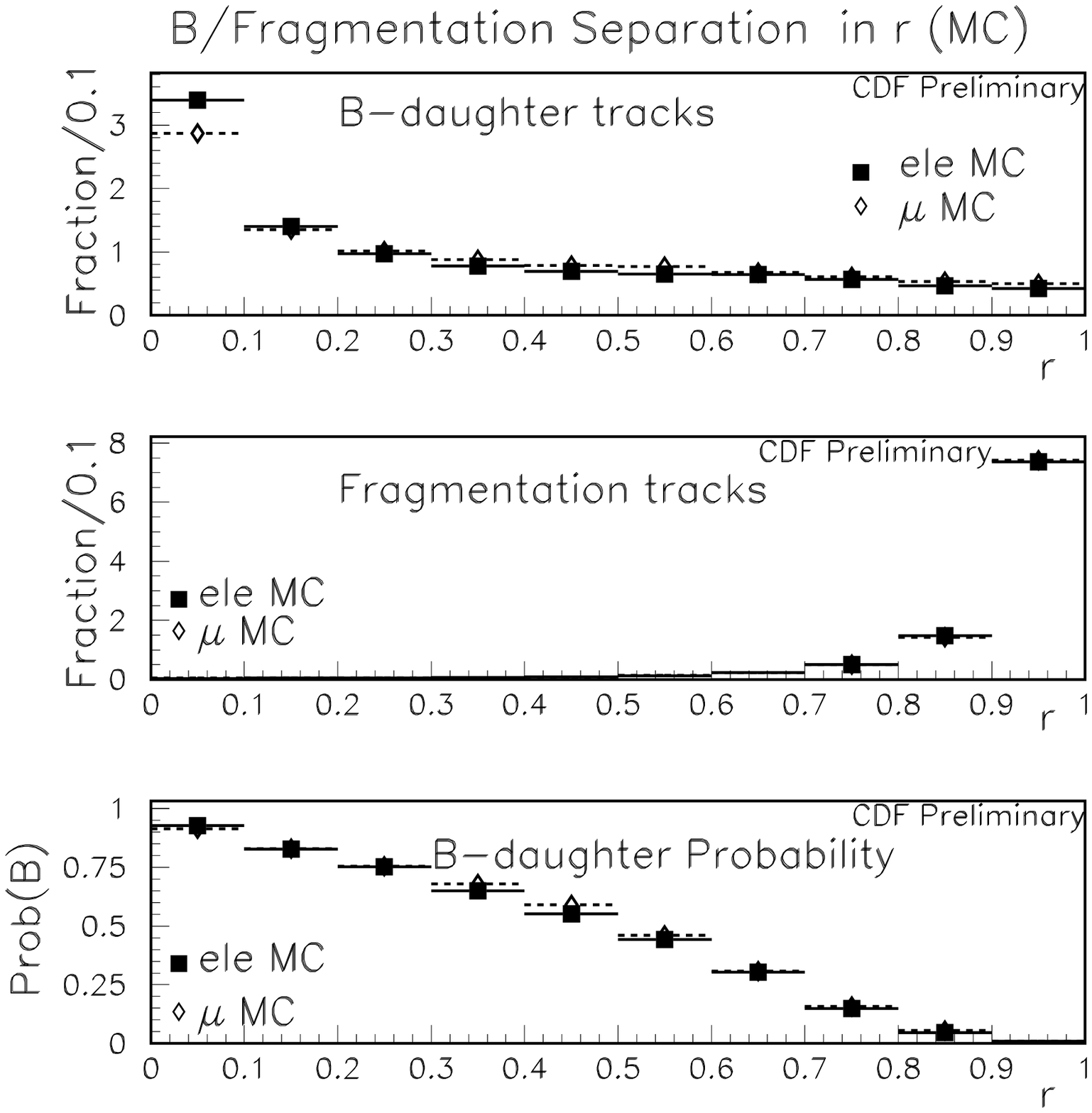}
  \epsfxsize=15pc 
  \epsfbox{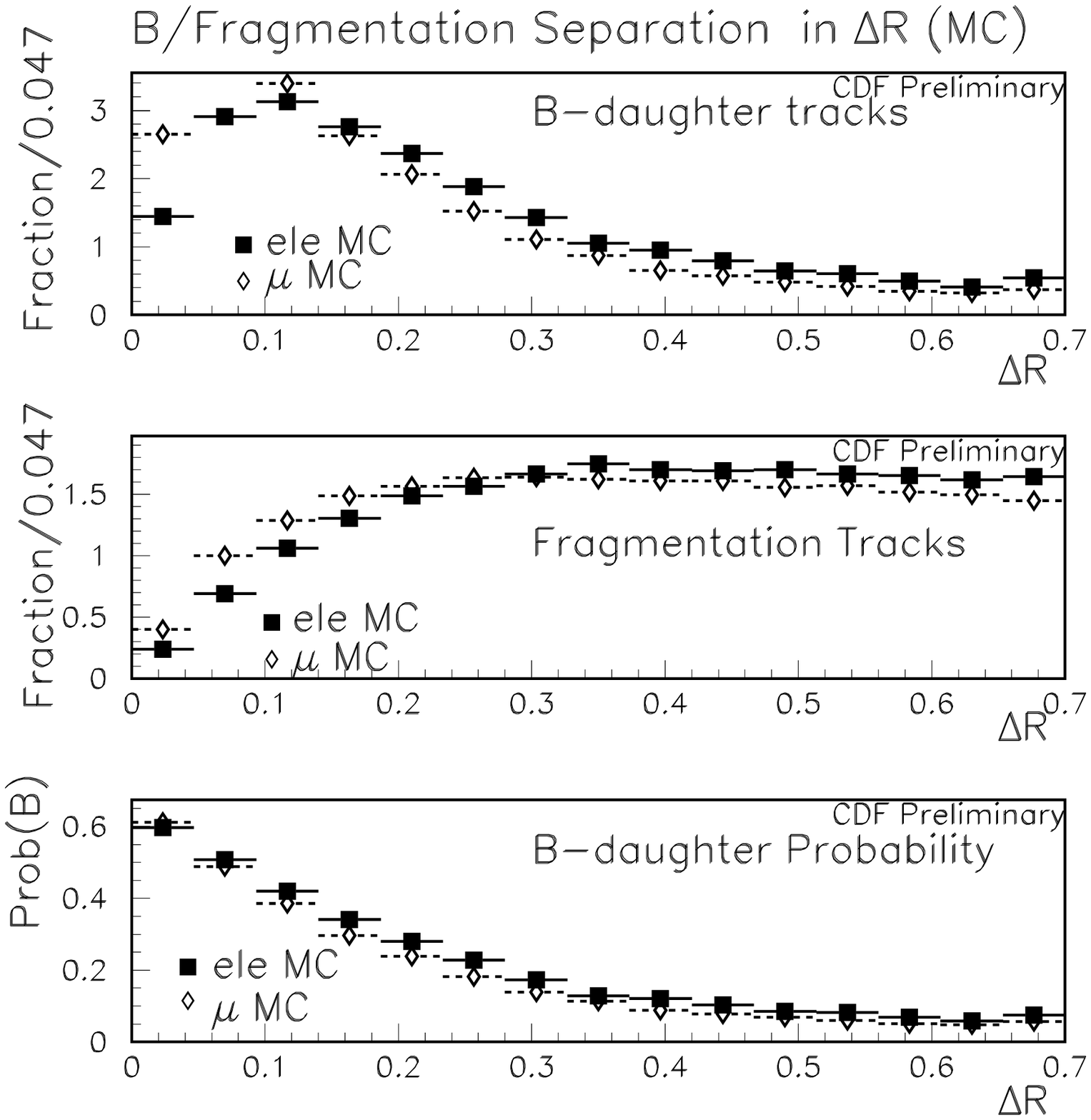}
}
%\vspace*{13pt}
%\vspace*{13pt}
\fcaption{The behavior of $B$-daughters (top),
fragmentation tracks (middle), and the
probability of being a $B$-daughter (bottom)
as calculated from Monte Carlo 
in the $r$ and $\Delta R$ variables.}
\label{fig:1a} 
\end{figure}

The cut on ${\cal P}_B(r,\Delta R)$
suppresses $B$ daughters as tag tracks,
and is defined as:
$$
{\cal P}_B(r,\Delta R) \equiv \frac{B(r,\Delta R)}
            {B(r,\Delta R) + F(r,\Delta R)},
\;\; {\rm with} \;\;
r \! =\!   \frac{d_{B}/\sigma_{B}}
           {d_{pv}/\sigma_{pv} + d_{B}/\sigma_{B} }
$$
where $d_{B}$ ($d_{pv}$) is the track impact parameter 
relative to the $B$ (primary) vertex,
$\sigma$ it's error,
and $\Delta R$ defined as usual.
$B(r,\Delta R)$ and $F(r,\Delta R)$ are the respective
numbers of $B$ and fragmentation tracks
in $r$ and $\Delta R$;
their behavior is shown in Fig.~\ref{fig:1a}, 
as well as the probability of being a $B$-track.
The function ${\cal P}_B(r,\Delta R)$
is the correlated 2-dimensional distribution, but
only the projections are shown in  Fig.~\ref{fig:1a}.

\begin{figure}[htbp]
%\vspace*{13pt}
\centerline{
  \epsfxsize=12pc 
  \epsfbox{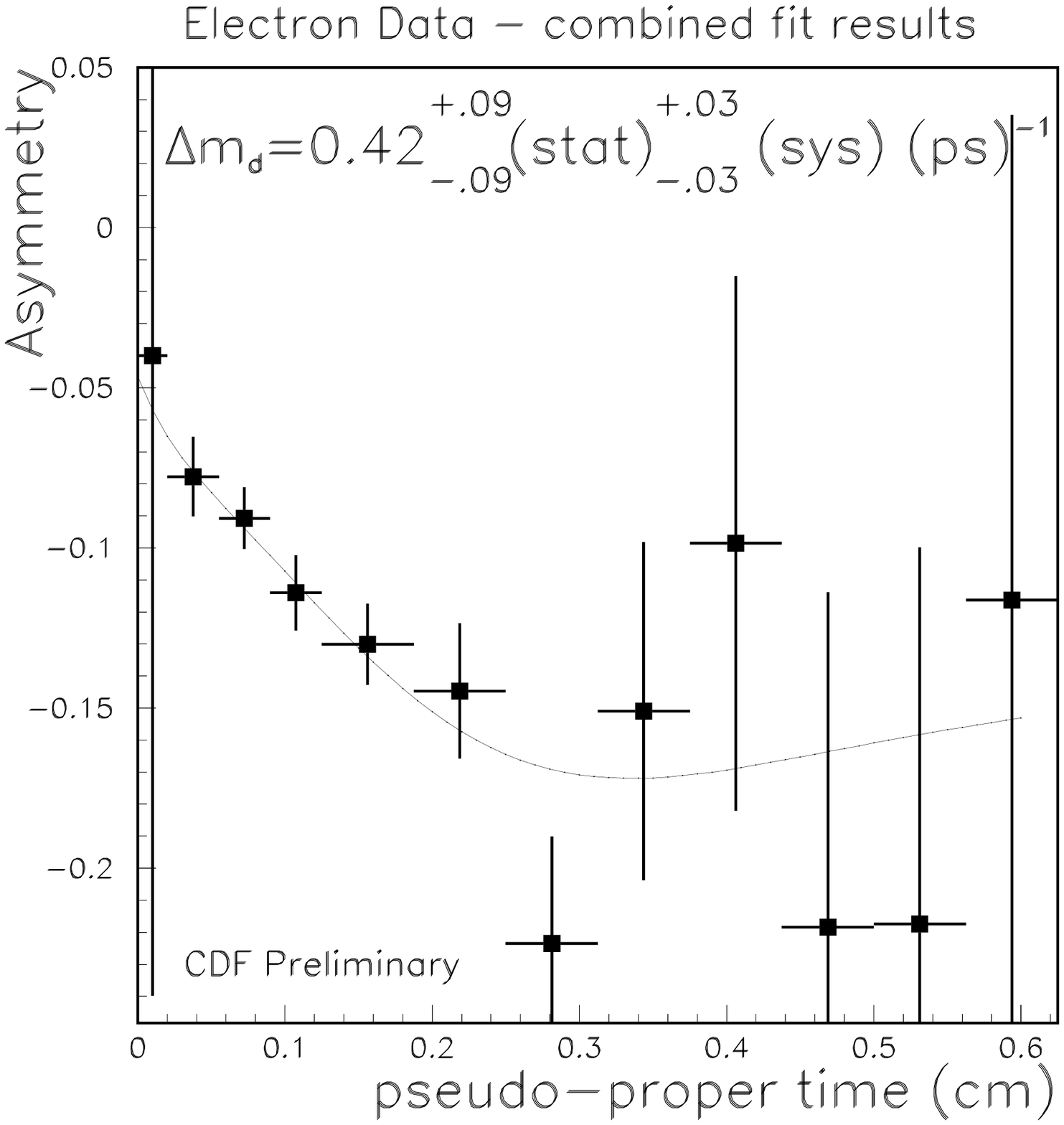}
  \epsfxsize=12pc 
  \epsfbox{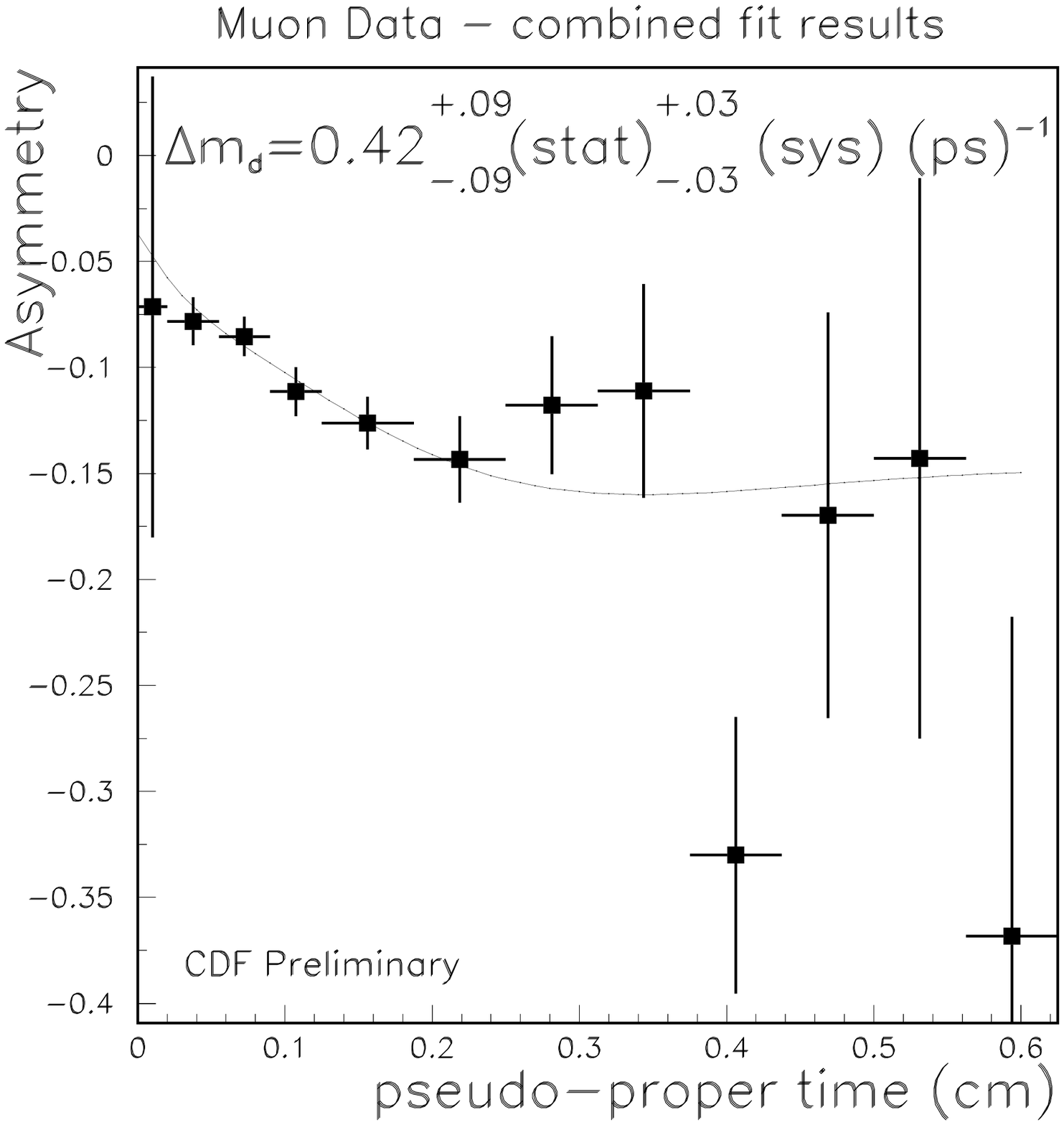}
% \vbox{\hrule width 5cm height0.001pt}
}
% \vspace*{1.4truein}		%ORIGINAL SIZE=1.6TRUEIN x 100% - 0.2TRUEIN
% \centerline{\vbox{\hrule width 5cm height0.001pt}}
\vspace*{13pt}
\fcaption{The raw asymmetry of SST tagged lepton$+$vertex data
(left: $e$; right $\mu$) fit to model of $B^0$ oscillations
and $B^+$, $B^0_s$, $\Lambda_b$, charm components.
The  $e$ and $\mu$ data are fit simultaneously
(note the different vertical scales).}
\label{fig:2} 
\end{figure}

\begin{figure}[tb]
% \vspace*{13pt}
% \vspace*{-5pt}
\centerline{
  \epsfxsize=19pc 
% \epsfbox{final.eps}
  \epsfbox{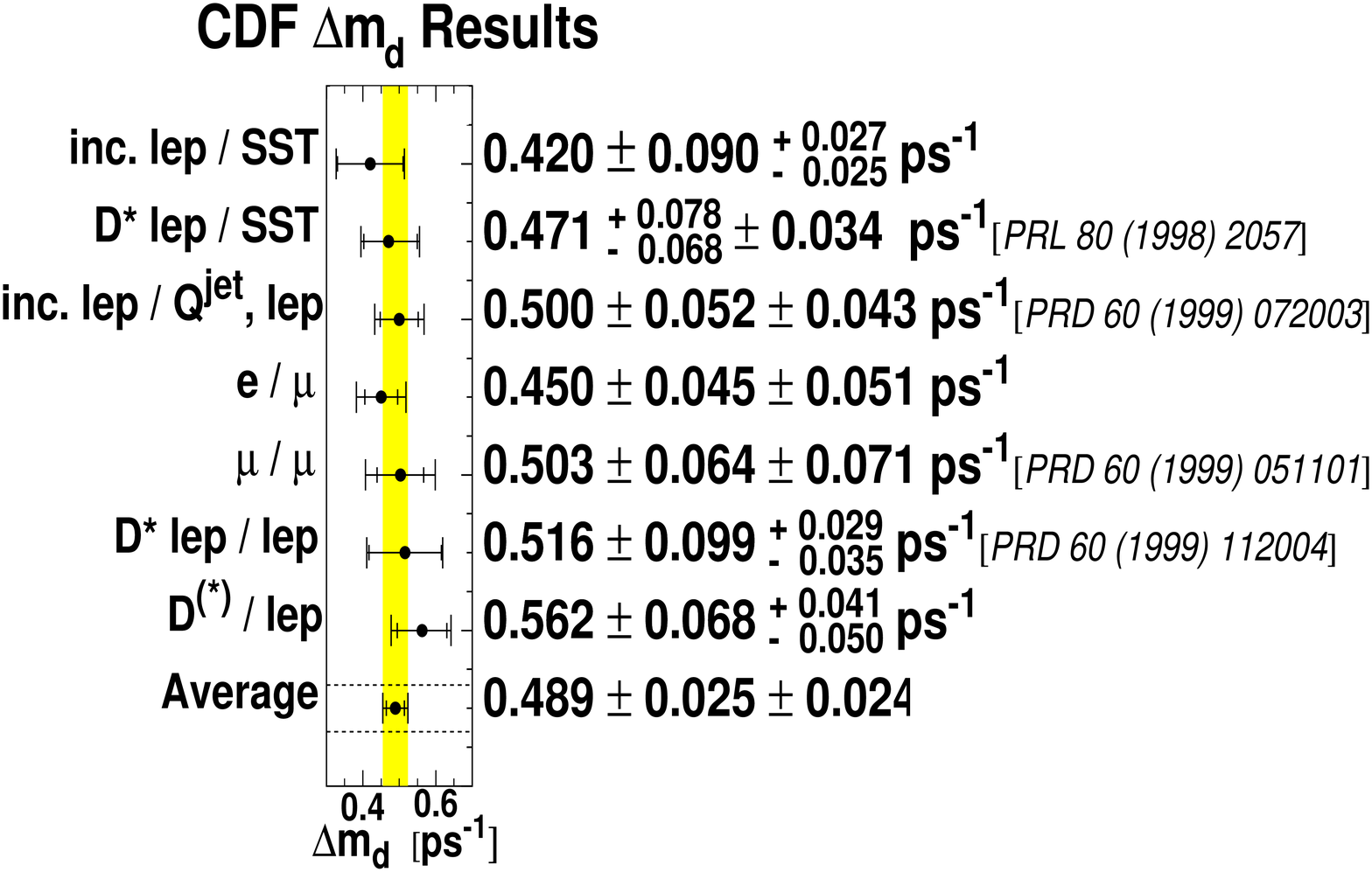}
  \epsfxsize=12pc 
  \epsfbox{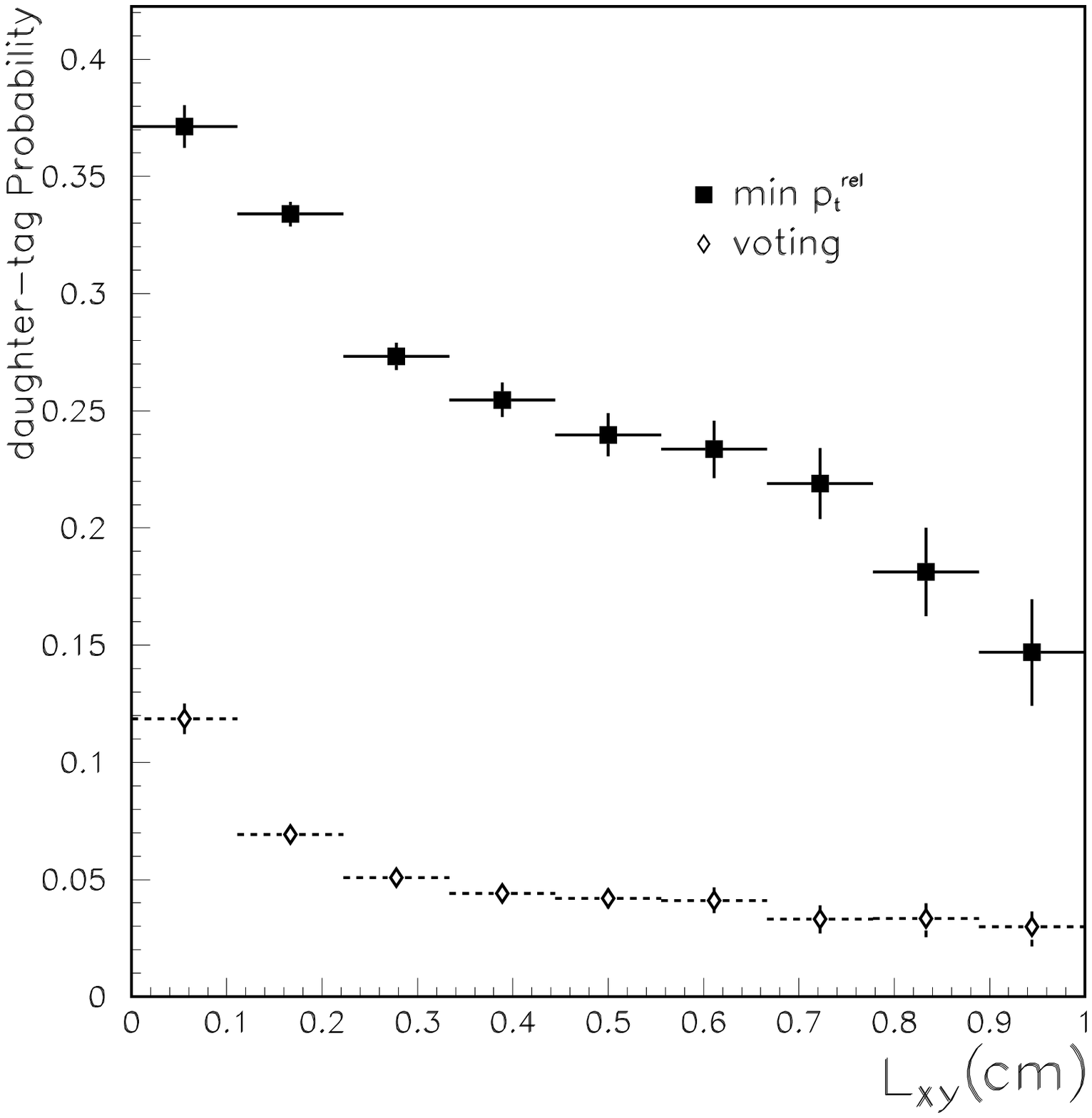}
}
\vspace*{13pt}
\fcaption{Left: Summary of CDF $\Delta m_d$ analyses.
          Right: Probability of tagging on a $B$-daughter
for the  $p_T^{rel}$ vs. probability SST in the $\ell+$vertex sample
as a function of transverse decay length.}
\label{fig:3} 
\end{figure}

With a tagged sample we compute the usual ``mixed''/``unmixed''
asymmetry ($\cal A$) in proper-time bins.
The apparent proper-time is corrected for lost decay pro\-ducts
by the usual methods (the ``k-factor'').
The raw asymmetry of the data is the sum of all sources;
that from $B^0_d$'s 
is ${\cal A}_0 = {\cal D}_0\cos \Delta m_d t$,
where ${\cal D}$ is the ``dilution'' 
({\it i.e.} $\!{\cal D} = 1-2P$, with mistag probability $P$).
Asymmetries due to 
$B^+$, $B^0_s$, $\Lambda_b$, and charm 
are included in the model, and while these have constant intrinsic 
asym\-me\-tries their fractional contributions or dilutions
can be time dependent.
This model is 
fit to the data using
a binned $\chi^2$, with
$\Delta m_d$, ${\cal D}(B^0)$, ${\cal D}(B^+)$,
and the fake-$\mu$ frac- 
tion 
free.
The result is shown in Fig.~\ref{fig:2},
which compares well with other single tag
CDF analyses in Fig.~\ref{fig:3}.
The $B^0$ dilution is found to be $13 \pm 3 ^{+2}_{-1}$\%, barely
$1\sigma$ small\-er
than in the ``$p_T^{rel}$'' SST applied to $\ell D^{(*)}$
events.\raisebox{-0.1ex}{\cite{SST}}
But, as seen in Fig.~\ref{fig:3}, this new ``voting''
method greatly suppresses tagging on $B$-daughters 
in $\ell+$vertex events.

%mix ref:
%e-mu:    1st conf proc, no pub???
% Time Dependent B0 anti-B0 Mixing at CDF 
% F. DeJongh, G. Michail, The CDF Collaboration, 
% FERMILAB-CONF-96/407-E. Published Proceedings 1196 Divisional
% Meeting of Division of Particles and Fields, American Physical Society, 
% Minneapolis, MN, August 10-15, 1996. 

% \begin{figure}[htbp]
% \vspace*{13pt}
% \centerline{\vbox{\hrule width 5cm height0.001pt}}
% \vspace*{1.4truein}		%ORIGINAL SIZE=1.6TRUEIN x 100% - 0.2TRUEIN
% \centerline{\vbox{\hrule width 5cm height0.001pt}}
% \vspace*{13pt}
% % \fcaption{Labeled tree {\footnotesize\it T}.}
% \end{figure}

%\vspace*{-10pt}
\nonumsection{References}

 \end{document}